\documentclass[figures,cite,debug]{epl}
\usepackage{amsmath,epsfig,latexsym,graphics,subfigure}
\newcommand{\be}{\begin{equation}}
\newcommand{\ee}{\end{equation}}
\newcommand{\bea}{\begin{eqnarray}}
\newcommand{\eea}{\end{eqnarray}}

\title{
Self-assembly of the gyroid cubic mesophase: lattice-Boltzmann
simulations}
\shorttitle{Lattice-Boltzmann simulations of gyroid phases}

\author{
N\'elido
Gonz\'alez-Segredo\thanks{n.gonzalez-segredo@ucl.ac.uk}
\and Peter V. Coveney\thanks{p.v.coveney@ucl.ac.uk}
}
\shortauthor{N. Gonz\'alez-Segredo \etal}

\institute{
Centre for Computational Science, Department of Chemistry, University
College London - 20 Gordon Street, London WC1H~0AJ, United Kingdom.
}


\begin{document}

\maketitle

\begin{abstract}
We present the first simulations of the self-assembly kinetics of the
gyroid cubic mesophase using a Boltzmann transport method. No
macroscopic parameters are included in the model and three-dimensional
hydrodynamics is emergent from the microscopic conservation laws. The
self-assembly arise from local inter-particle interactions in an
initially homogeneous, phase segregating binary fluid with dispersed
amphiphile. The mixture evolves in discrete time according to the
dynamics of a set of coupled Boltzmann-BGK equations on a lattice. We
observe a transient microemulsion phase during self-assembly, the
structure function peaks and direct-space imaging unequivocally
identifying the gyroid at later times. For larger lattices, highly
ordered subdomains are separated by grain boundaries. Relaxation
towards the ordered equilibrium structure is very slow compared to the
diffusive and microemulsion-assembling transients, the structure
function oscillating in time due to a combination of Marangoni effects
and long-time-scale defect dynamics. 
\end{abstract}

Block copolymer melts or dispersions, and homopolymer-block copolymer
blends are examples of systems that self-assemble into regular,
liquid-crystalline structures when subjected to the appropriate
temperature or pressure quenches
\cite{REVIEW1,REVIEW2,LAURER,HAJDUK}. These structures, called  
mesophases due to their features being intermediate between those of a
solid and a liquid, are also found in fluid mixtures of a surfactant
in a solvent, binary immiscible fluids containing a third, 
amphiphilic phase, and lipidic biological systems \cite{REVIEW1,BIO}. 
They all form due to the competing attraction-repulsion mechanism
between the species. The morphology of these mesophases is defined by
the spatial loci where most of the amphiphile concentrates, forming
multi- or mono-layer sheets of self-assembled amphiphile.  Common
equilibrium mesophases include lamellae, hexagonal columnar arrays,
and the primitive ``P'', diamond ``D'' and gyroid ``G'' cubic phases
\cite{REVIEW1,REVIEW2}. The sheets of these cubic phases are
surfaces or {\em labyrinths} of zero mean curvature, the skeletons of
which form double (inter-weaving), chirally symmetric bicontinuous cubic
lattices which are 6-, 4- and 3-fold coordinated, respectively. The
gyroid is the phase which exhibits the least surface area per unit
cell among those, and is ubiquitous in nature; we show in this paper
that it can spontaneously self-assemble from a purely microscopic,
kinetic-theoretical lattice model with hydrodynamic interactions.

Simulation approaches to the dynamics of mesophase formation 
have been hitherto based on Monte Carlo \cite{MC}, Brownian
dynamics \cite{GROOT2}, dissipative particle dynamics
\cite{GROOT1,GROOT2,PRINSEN}, Langevin diffusion equation 
\cite{NONOMURA_IMAI_QI,ZVELINDOVSKY_VAN_VLIMM} and molecular dynamics
methods \cite{MARRINK}. In Langevin-diffusion methods, mass currents
arise from chemical potential gradients, computed in turn from
equilibrium free energies. Much of the published work is based on 
Ginzburg-Landau expansions for the latter, assuming that all
surfactant is adsorbed as a continuum on the self-assembled sheets,
incorporating white noise and excluding hydrodynamics
\cite{NONOMURA_IMAI_QI}. More recently, extensions appeared including
hydrodynamics and free energies explicitly calculated for the
amphiphile, modelled as Gaussian bead-spring chains in a mean-field
environment \cite{ZVELINDOVSKY_VAN_VLIMM}. Dissipative particle
dynamics (DPD) methods also model the amphiphile as bead-spring
chains, yet the beads as well as the particles constituting 
the fluid species enter in the model as mesoscopic entities, undergoing
2-body interactions. In DPD, space is continuous and hydrodynamics is
emergent from the mesodynamics. The presence of hydrodynamics is an
important feature in modelling the nonequilibrium pathways of
mesophase self-assembly and the possible metastable states they can
lead to, but is absent in Monte Carlo \cite{MC} and Brownian dynamics
\cite{GROOT2} methods.  

In this work we use a hydrodynamically correct lattice-Boltzmann model
of amphiphilic fluids \cite{CHEN_NEKOVEE} to simulate the
self-assembly of a liquid crystalline, double gyroid cubic phase from
a randomly mixed initial binary immiscible fluid (say, of ``oil'' or
``red'', and ``water'' or ``blue'') with an amphiphilic species
dispersed in it. The model \cite{CHEN_NEKOVEE} does not
require the existence of a thermodynamic potential describing the
local equilibria and a  phase transition; rather, self-assembly arises
as an emergent property of the microscopic interactions between the
species. The dynamics is obtained by solving a set of coupled
Boltzmann-BGK transport equations on a spatial lattice in discrete
time steps with a discrete set of microscopic velocities; the scheme
is known as the lattice-Boltzmann (LB) BGK method and has proved
useful for single- and multi-phase flow modelling during the last
decade \cite{SUCCI}. At each time step, the probability density
evolved by each LB equation is advected to nearest neighbours and
modified by molecular collisions, which are local and conserve mass
and momentum. A single time parameter controls relaxation in the
collision term for all 
microscopic speeds, and in our model there is no stochastic noise
present other than in the amphiphile dynamics. The mass density
defines fluid elements on each lattice node which can be mapped onto 
experimental scales that are intermediate with respect to molecular
and macroscopic lengths and times. Immiscible fluid behaviour is 
incorporated via scalar inter-particle forces of a mean-field form
limited to nearest neighbours. The force enters in the hydrodynamics
by modifying the local macroscopic velocity of the whole fluid and
hence the local Maxwellian to which each species relaxes. For the
correct lattice symmetry, and in the limits of low Mach and Knudsen
numbers, the Navier-Stokes equations for incompressible flow hold in
the bulk of each fluid phase. The model also leads to the growth
exponents and dynamical self-similarity observed in binary immiscible
spinodal decomposition experiments \cite{P2}. An amphiphile density is
evolved by an additional coupled LB equation, and the bipolar,
amphiphilic molecules are modelled as dipole vectors moving between
the nodes of the lattice. Their orientations vary continuously and
relax towards a Gibbsian canonical equilibrium which minimises the
interaction energy between the local dipole and the mean fields
generated by their nearest neighbours. The evolution of the surfactant
density is also coupled to that of the other species
\cite{CHEN_NEKOVEE}. 

\begin{figure}[!htb]
\begin{center}
\includegraphics[angle=0,height=8cm]{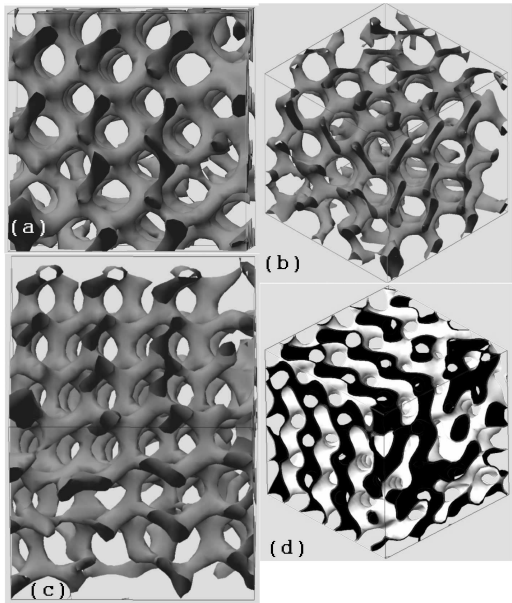}
\caption{\small Isosurfaces of the order parameter
$\phi(\mathbf{x})$ for a surfactant density of
$n^{\mathrm{(0)s}}=0.60$ at time step $t=15000$ in a highly ordered
$33^3$ subdomain of a $128^3$ lattice. Panels (a), (b), and  (c)
display the $\phi=0.40$, high-density isosurface viewed along the
$(100)$, $(1\,\overline{1}\,\overline{1})$ and $(110)$ directions,
respectively. Panel (d) shows the oil-water interface ($\phi=0$) of
the same lattice subdomain along the direction
$(1\,\overline{1}\,\overline{1})$, where between three and four units
cells fit laterally in the subdomain. Black and white have been used
in panel (d) to distinguish one immiscible fluid phase from the other,
and the scale for panels (a) and (c) varies from that for (b) and (d).
\normalsize}     
\label{ORDER_PARAM}
\end{center}
\end{figure}

The initial condition in our simulations is a random dispersion of
surfactant in a random mixture of equal amounts of oil and water. The
maximum values chosen for the (uniformly distributed) random densities
were 0.70 for oil or water and $0.40, 0.60,
0.90\equiv~n^{\mathrm{(0)s}}$ for the surfactant. These, as   
well as all magnitudes and parameters in the model, are in lattice
units. All relaxation times were set to 1.0, the thermal noise
parameter for dipolar relaxation was set to 10.0, and the parameters
controlling the strength of the inter-species forces were set to
$g_{\mathrm{br}}=0.08$, $g_{\mathrm{bs}}=-0.006$ and
$g_{\mathrm{br}}=-0.003$. We employed periodic boundary conditions on
cubic lattices of $64^3$, $128^3$ and $256^3$ nodes, the latter two
initially being required to check that finite size effects were
absent. The choices made of densities and parameters were based on
previous tests searching for regimes of oil-water immiscibility ({\em
  i.e.}, below the spinodal) for which, within the computing time and
resources available, phase segregation was sufficiently fast while
flows were dominated by hydrodynamics and surface tension as opposed
to diffusion.  

We are interested in studying the nonequilibrium pathways that follow
from the initial condition, for which we track the evolution of mass
densities via direct-space imaging and analysis of the structure
function. Defining a scalar order parameter, $\phi(\mathbf{x})$, at 
a particular time step as the oil density minus the water density, the
oil-water structure function, $S(\mathbf{k})$, is the Fourier
transform of the spatial auto-correlation function for the
fluctuations of $\phi(\mathbf{x})$ around its lattice average,
proportional to the intensities probed in scattering techniques widely
used in the analysis of mesophase structure; the spherically averaged
structure function, $S(k)$, is the average of $S(\mathbf{k})$ 
in a shell of radius $k\equiv|\mathbf{k}|$ and thickness one lattice 
unit, corresponding to the contribution of structures of size
$L\equiv2\pi/k$.  

Figure \ref{ORDER_PARAM} shows isosurfaces of the order parameter at
time step $t=15\,000$ in a $33^3$ subdomain of a $128^3$ lattice for an
initial surfactant density flatly distributed up to
$n^{\mathrm{(0)s}}=0.60$. We display three viewpoints of the
isosurface $\phi=0.40$ (in lattice units), corresponding to a
water-in-oil, ``rod-like'' scenario where water is a minority phase
and oil is in excess. Whereas on $64^3$ (or smaller) lattices the
liquid crystalline structure uniformly pervades the simulation cell,
on $128^3$ (or larger) lattices there are some imperfections present,
more prevalent as the lattice size is increased, resulting in liquid
crystalline subdomains with slightly varying orientations between
which exist domain boundaries---``defects". 

\begin{figure}[!htb]
\begin{center}
\twoimages[angle=0,width=5.5cm]{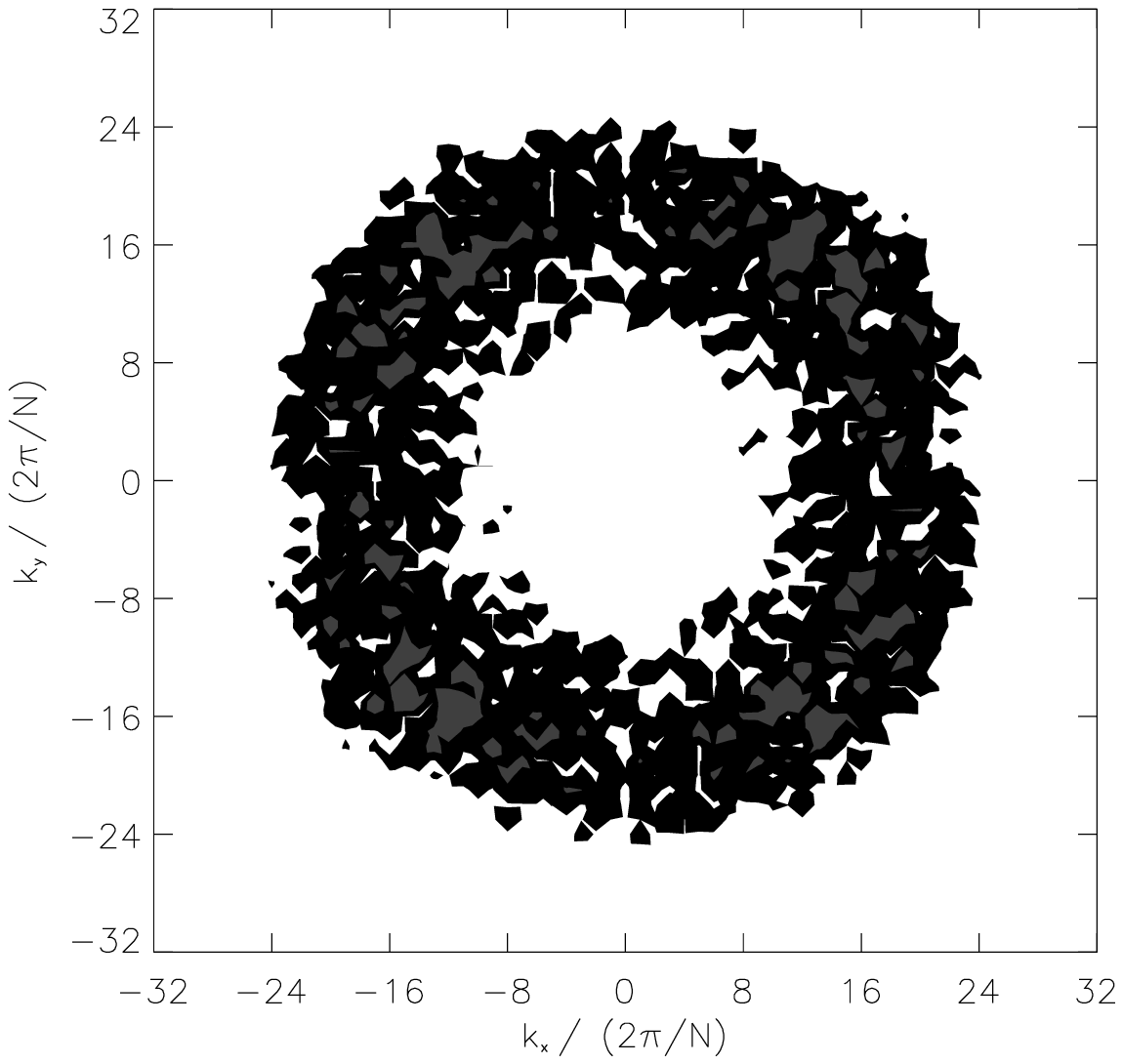}{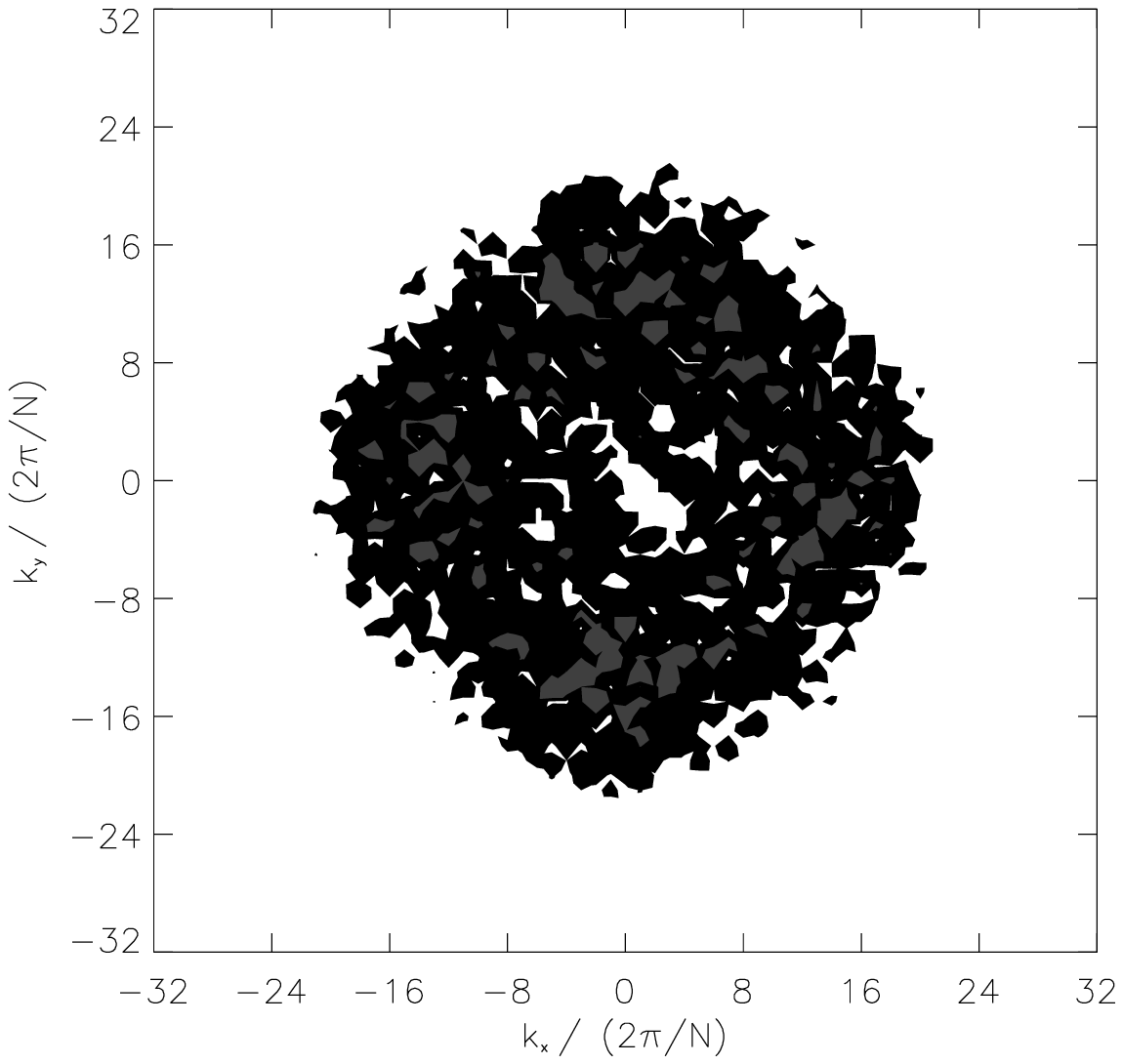}
\twoimages[angle=0,width=5.5cm]{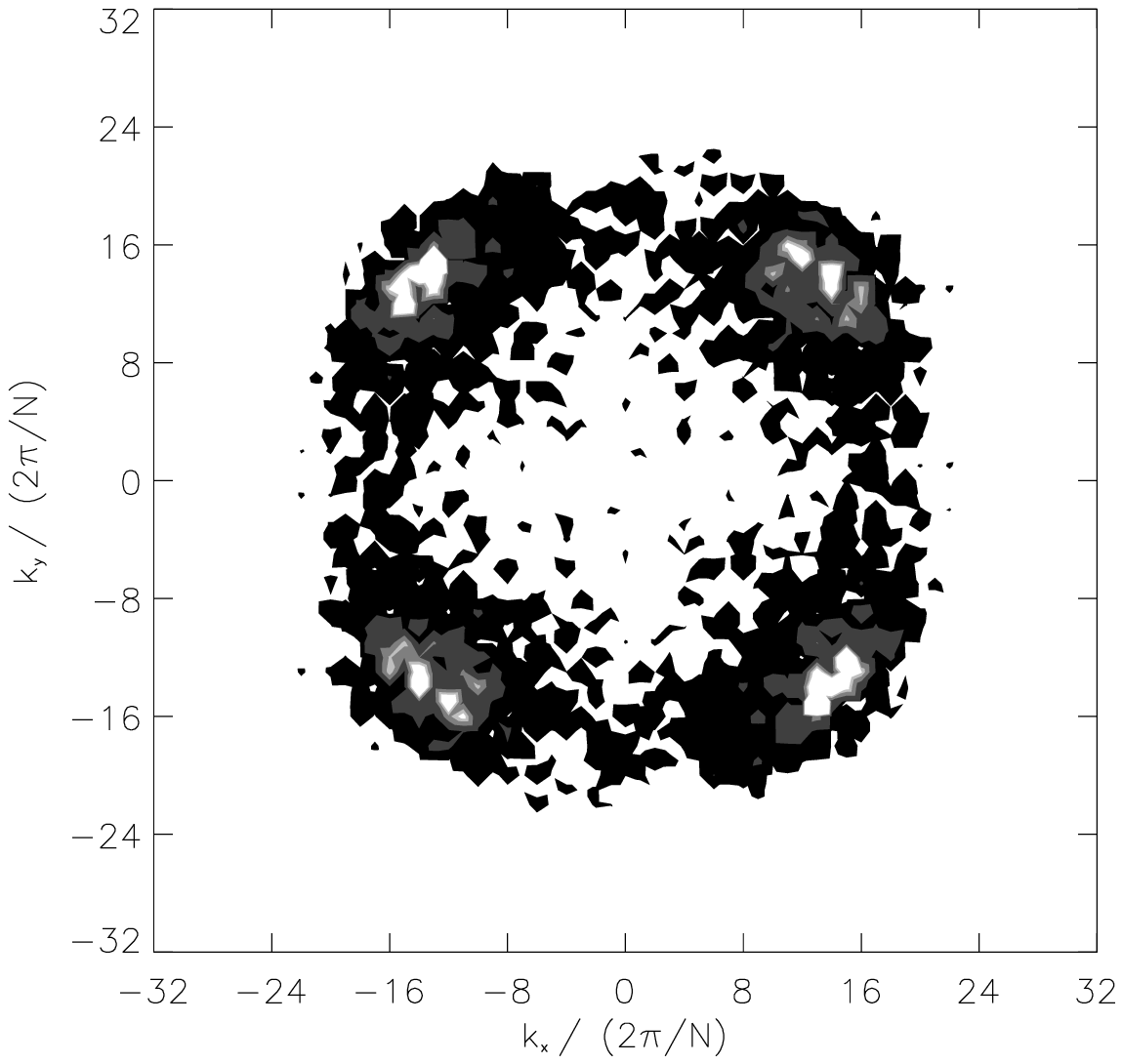}{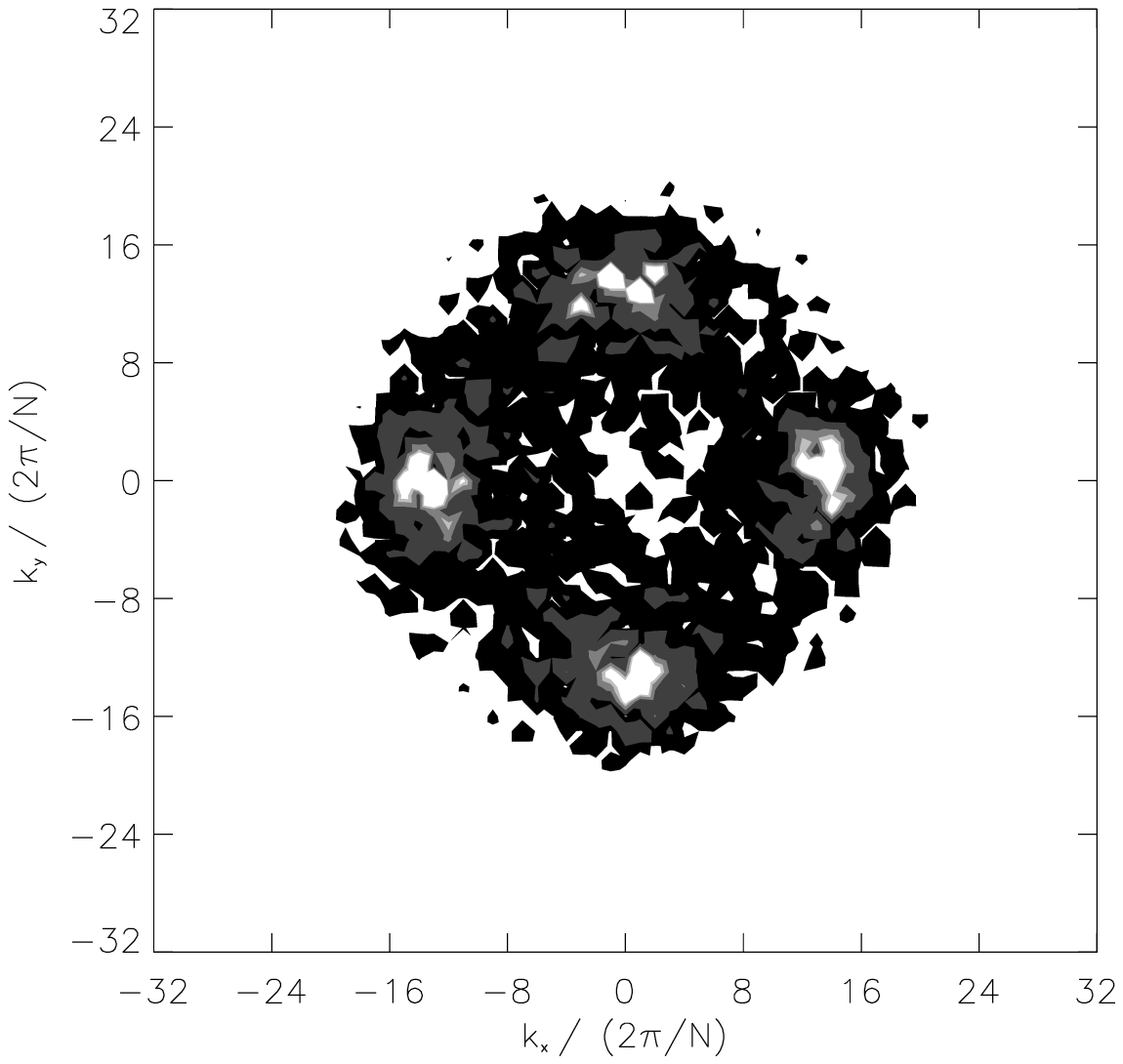}
\caption{Temporal evolution of $(k_x,k_y)$ slices of the structure
function viewed along the $(100)$ direction. The fluid is the same of
Fig.~\ref{ORDER_PARAM}. Left and right columns 
show slices at $k_z/(2\pi/N)=0,\pm14$, respectively, where $N=128$ is
the lattice lateral size. Rows from top to bottom correspond to time
steps $t=500$ and $15000$, respectively. Contours denote intensities
$S=1,50$ and $100$, where lighter shades denote higher
intensities. The spherical shell structures in the top row indicate
the presence of a sponge (microemulsion) phase, which becomes
anisotropic at later times. In lattice units. 
\normalsize}     
\label{SF3D}
\end{center}
\end{figure}

The resemblance of the simulated structures in Fig.~\ref{ORDER_PARAM}
to transmission electron microtomography (EMT) images of the gyroid
``G" cubic morphology is evident \cite{HAJDUK}: the morphology of
$\pm\phi,\,\phi\ne0$ (excess) isosurfaces is that of gyroid
skeletons. The lattice resolution 
is insufficient to detect multiple peak fingerprints in plots of
$S(k)$, as observed experimentally with SAXS techniques
\cite{LAURER,HAJDUK}. However, its unaveraged counterpart
(Fig.~\ref{SF3D}) shows complete agreement of ratios of reciprocal 
vector moduli with those observed in diffraction patterns of the
gyroid \cite{HAJDUK}, which, in addition to visual (direct space)
inspection of the unit cell, leads to unequivocal identification. 
EMT images and experimental self-assembly times of the gyroid phase
allows us to broadly associate a length and time scale to our LB
dynamics; {\em e.g.} the systems in \cite{HAJDUK,SQUIRES} require
resolutions to be 2.3~nm per LB-lattice unit and $10~\mu{}$s up to
$40~$ms per LB time step. 

Values higher than $n^{\mathrm{(0)s}}=0.60$ also produce gyroid
structures at late times, whereas there is gradual loss of long-range
ordering for $0.40<n^{\mathrm{(0)s}}<0.60$, leading to a molten gyroid
phase. At $n^{\mathrm{(0)s}}=0.40$, the late-time structure becomes a
{\em~sponge} (microemulsion), isotropic and of short-range order, for
which $S(\mathbf{k})$ is similar to that shown in Fig.~\ref{SF3D}, top
row. Although the observation of gyroid-related morphologies has
earlier been claimed in
Langevin-diffusion~\cite{ZVELINDOVSKY_VAN_VLIMM} and DPD~\cite{GROOT1}
methods, the evidence was purely pictorial and not comprehensive---in
fact, the structures more closely resemble molten gyroid states.  

\begin{figure}[!htb]
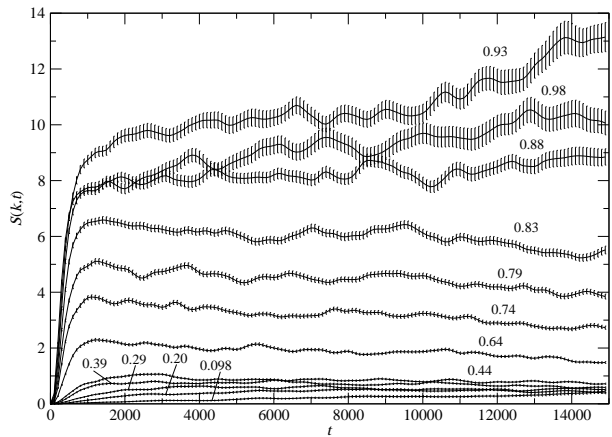

\onefigure[angle=0,width=8cm]{eps/figure3.eps}
\caption{Temporal evolution of the spherically averaged structure
function and its error (one standard deviation uncertainty in the
average), for the wavenumbers indicated next to each curve. The fluid
is the same as in Fig.~\ref{ORDER_PARAM}. All quantities are in
lattice units.}       
\label{SF_TEMPORAL}
\end{figure}

Figure \ref{SF_TEMPORAL} shows the time evolution of the spherically
averaged structure function for the $n^{\mathrm{(0)s}}=0.60$ gyroid
case and wavenumbers corresponding to average domain sizes in
$6.4<L<64$. Note three characteristic features of the curves: there
are oscillations, decay (for $k<0.83$ and $k>1.1$, the latter not
shown), and growth (for $0.88<k<1.0$ or $6.1<L<7.1$, close to the
average domain size value). Modes of $k>1.3$ ($L<4.7$) decay fast
enough ($S(k)<0.1$ for $t\approx1000$) that they do not contribute to
the structure. The analogue curves for the $n^{\mathrm{(0)s}}=0.40$
case, leading to a sponge phase, exhibit similar oscillations, albeit
of longer period. The fact that there are two types of temporal
evolution, corresponding to increasing and decreasing modes, is a
reflection of a phase segregation process still taking place. In the
course of time, more domains accumulate in the $0.88<k<1.0$ range; we
discard increasing interface steepness as another contributing factor
to this since oil/water diffusion is negligible in this regime.

Direct-space observation of the interface ($\phi=0$) superimposed on
surfactant density maps for a series of time slices allows us to
ascribe the oscillations that mainly affect the growing modes to
``self-sustained'' Marangoni effects. These are caused by collective
and inhomogeneous amphiphile adsorption and desorption to and from
the (periodically modulated) interface, a region of high surfactant
density. This permits us to set a time scale of between 100 to 500
time steps during which interfacial surfactant from regions of high
density diffuses towards an adjacent interface (also with adsorbed
surfactant), effectively forming a bridge between the two 
sheets. If the interfaces belong to boundary, inter-domain regions,
where the global translational symmetry is broken, defects form,
change shape and annihilate on the same time scale. Perusal 
of Fig.~\ref{SF3D} confirms that the smallest period of the
oscillations is not dissimilar to such a time scale. The frequency
spectrum of the time evolution of $S(k)$ gives a rich structure of
peaks with a decaying envelope, higher-frequency modes becoming
excited as the surfactant concentration increases.

Direct-space observation also shows an essential feature of the
mesophase dynamics: each unit cell in the gyroid wanders in time
about a fixed spatial position, whereas for the sponge the interface
shows nonperiodic displacements. In other words, the temporal average
of the displacement is zero for a gyroid's unit cell and non-zero for
an interface element in the sponge phase. Displacements are small in
the former: they are not larger than {\em ca.} 20\% of the unit cell
size and are in-phase with those of the unit cells belonging to the
local, defect-delimited subdomain. We therefore expect the gyroid
structure to be stable, {\em i.e.} the dynamics would relax to it for
late times. 

Temporal oscillations in the structure function of sponge phases have
been reported previously by Gompper \& Hennes via a stochastic
Langevin diffusion equation method with hydrodynamics, based on a 
Ginzburg-Landau free energy \cite{GOMPPER}. This approach does not
explicitly consider an order parameter for the amphiphile since it
assumes that amphiphile relaxation is fast compared to that of the
oil-water order parameter. The oscillations reported therein range
from overdamped to 
underdamped depending on the wavenumber, and their frequency spectrum
shows a single peak at finite frequency. This is in contradistinction
to our finding of multiple-peak spectra, which we ascribe to the
absence in that model of scalar or vector degrees of freedom for the
amphiphile. In fact, Gompper~\&~Hennes put forward a linearised
Navier-Stokes model for Poiseuille flow wherein oscillations arise due
to incompressibility competing with pressure gradients. Our LB method
reproduces the same linearised, incompressible Navier-Stokes dynamics
away from interfaces for the quiescent flows we observe, and, despite
this, we obtain multiple-peak spectra. In addition, since stochastic
sources are not present in the oil/water evolution, they cannot
account for this spectral multiplicity. While it is true that
randomness in the adsorbed surfactant directors may effectively reduce
amphiphile adsorption strength, this effect is negligible compared to
surfactant diffusion currents, facilitated by gradients of
$\phi(\mathbf{x})$ from nearby interfaces.  

Since the systems we simulate are dissipative and isolated (there is
no mass or momentum exchange with external sources), oscillations are
bound to die out at sufficiently late times. We observe interfacial
widths to have 
reached their minimum (and hence interfacial tension its maximum)
at about time step 1000, at which time the structure has a sponge-like
morphology. Then the structure undergoes slow relaxation on a time
scale which is $\mathcal{O}(10^4)$. We observe the pathway to
equilibration to be a slow process dominated by currents created by
surface tension and Marangoni effects acting on similar time scales. A
free energy ``leading the way'' towards the equilibrium morphology
might be less useful than a correct mesodynamics, and even bias the
evolution; methods which are intrinsically mesoscopic
{\em~bottom-up} such as ours and DPD are better suited. From DPD
simulations of copolymer melts \cite{GROOT1}, Groot \& Madden found
that melts of symmetric amphiphile led to lamellar phases, whereas a
gyroid-like structure appeared only for asymmetric amphiphile as a
transient phase precursor to a hexagonal columnar phase. Our
results are in contradistinction to these: although our amphiphiles
are symmetric, the gyroids we find are stable.  

Finally, our reproduction of periodically modulated mesostructures
rebuts claims that a necessary condition for their self-assembly is a
disparity in the ranges of interaction of the competing morphogenic
mechanisms, namely, short range versus long range \cite{SEUL}. Unlike
other mesoscopic approaches such as DPD, ours is strictly local, being
based on nearest neighbour interactions on a lattice.

The simulation of the gyroid cubic phase reported here highlights the
richness of our model's parameter space. Our LB model represents a
kinetically and hydrodynamically correct, bottom-up, mesoscale
description of the generic behavior of amphiphilic fluids, which is
also algorithmically simple and extremely computationally efficient on
massively  parallel platforms \cite{LOVE}. Natural extensions of this
work include the search for regimes leading to equilibrium mesophases of
varied symmetries, the study of shear-induced symmetry transitions and
the analysis of defect dynamics in large scale simulations. 
Computational steering tools should prove invaluable in these respects
\cite{STEER}. 

\acknowledgments

Access to supercomputer resources was 
provided by the UK EPSRC under grants GR/M56234 and 
RealityGrid GR/R67699.

\end{document}